\documentclass[conference]{IEEEtran}
\usepackage{cite}

\usepackage[pdftex]{graphicx}
\graphicspath{{./figures/}}

\usepackage{amsmath}
\usepackage{algorithmic}

\usepackage{textcomp}
\usepackage{xcolor}
\def\BibTeX{{\rm B\kern-.05em{\sc i\kern-.025em b}\kern-.08em
    T\kern-.1667em\lower.7ex\hbox{E}\kern-.125emX}}

\usepackage{listings, multicol}
\usepackage{enumitem}
\usepackage[export]{adjustbox}
\usepackage{fancyvrb}

\usepackage{multicol}

\begin{document}
\def \DSLname{\emph{SWIG}} 
\def \FWname{\emph{ODESSA}}
\def \DTname{\emph{YYY} } 

\title{Towards Implementing ML-Based Failure Detectors}
%


\author{\IEEEauthorblockN{Xiaonan Li and Olivier Marin}
\IEEEauthorblockA{New York University Shanghai, Shanghai, China\\
emails: [xl2149, ogm2]@nyu.edu}}

\maketitle              
\begin{abstract}
  Most existing failure detection algorithms rely on statistical methods, and very few use machine learning (ML). This paper explores the viability of ML in the field of failure detection: is it possible to implement an ML-based detector that achieves a satisfactory quality of service?
  We implement a prototype that uses a basic long short-term memory neural network algorithm, and study its behavior with real traces. Although ML model has comparatively longer computing time, our prototype performs well in terms of accuracy and detection time.
  
\end{abstract}
\begin{IEEEkeywords}
  failure detection, machine learning.
\end{IEEEkeywords}

  \vspace*{-1mm}

\section{Introduction}


Distributed systems should provide reliable and continuous service despite  failures. 
In~\cite{chandra_toueg_1996}, Chandra and Toueg show that failure detection is the dominant factor in system unavailability, and
introduce the notion of unreliable failure detector (FD) as a theoretical construct to extend the applicability of distributed algorithms such as consensus and atomic broadcast. An FD is an oracle which monitors a remote process $P$, and assesses in real time whether $P$ is up or has crashed. The assessment is unreliable because an FD might provide incorrect information over limited periods of time. 

Machine learning (ML) performs well upon analyzing extremely regular data, but network latency on a link can vary often and significantly over time, and failures or delays are transient events that occur irregularly. To achieve high accuracy rates consistently, an ML-based FD must constantly train over newly emerging data. Resource greediness can be prohibitive, since a crucial aspect of an FD is its output rate. Our main goal is to study the viability of FDs based on ML techniques; such FDs should require minimum resources and time during real-time training while maintaining high accuracy. In this paper, we present a preliminary approach that uses a basic long short-term memory neural network algorithm. After comparing its output with a baseline FD, we further optimize our model’s performance by adjusting its parameters and structure. Our simulations based on real traces show that our model performs well in terms of accuracy and detection time, despite that the ML model incurs non-negligible computation time.

\section{Related work}
\label{related}

An FD is a process that monitors remote processes, and strives to estimate whether they're still up or whether they've crashed. The authors of~\cite{FLP85} prove formally that there is no way to determine the failure of a node in a distributed environment with 100\% certainty. To make up for this, distributed systems can establish a network of mutual observation via \emph{unreliable} FDs~\cite{chandra_toueg_1996}. Such FDs require each node in the system to send periodical heartbeat messages to a monitor node to prove its liveliness. If a heartbeat message does not arrive before its expected arrival date, the monitor will suspect that the corresponding process has failed. A high performance FD is expected to generate a series of efficient but tolerant expected arrival dates, which are able to detect a failed process in time with minimum wrong suspicions of the healthy ones.

Implementing an FD is a challenge, because it must contend with the unpredictable and asynchronous nature of network links while preserving its set properties in terms of completeness and accuracy~\cite{MYBELOVED_PROFESSORS_PAPER_LETS_CITE_IT}. Early approaches~\cite{Fetzer_addaptive_2001,sotoma_2001} propose adaptive FDs that adjust the timeout delay tolerance dynamically; but they assume an unrealistic timing model with no bound on the delays. Chen et al.~\cite{chen_toueg_aguilera_2002} overcome this issue by introducing quality of service metrics which quantifies the performance of an FD based on the speed of correct detection and the ratio of a false ones.




Previous FDs are mostly based on statistical models. Although more recent studies introduce ML-driven methods for FD-related problems, their solutions are developed under different context. In \cite{log_anomaly_detection}, the authors use a stacked-LSTM model to classify potential anomaly events in cloud services by analyzing labeled static sensor logs. They conclude that their S-LSTM approach has the ability to quickly learn from the historic patterns and adjust to unpredictable anomalous events. The authors of~\cite{fd_sci_ds} propose three different ML methods for high-performance computing systems failure detection by analyzing informative hardware usage data. Their best solution, based on the Support Vector Machine method, achieves a precision of 90\%. \cite{system_fd_dl}~acknowledges the significance of timestamp data in system logs, and passes both time and token sequence data into the model under the form of interpolated vectors. By using a CNN model, this approach reaches a 99.5\% F1 score. 

The ML-driven methods mentioned above address a different, extended model where they might have more extensive information for error implication and require pre-labeled data to tag anomalies in the original datasets, addressing a classification problem. Because of this, they don't apply well to the classic FD problem which is a regression problem since timestamps are the only retrievable information from the monitored sub-systems, yet this information is self-sufficient for detection training. Hence, in this paper we will focus on studying the viability of ML on the classic FD problem and compare the results with the statistic-driven FD models.

\section{System Model}
\label{set_up}

\begin{figure}[htb]
\vspace*{-3mm}
\hspace*{-0.2cm}
\includegraphics[scale=0.164]{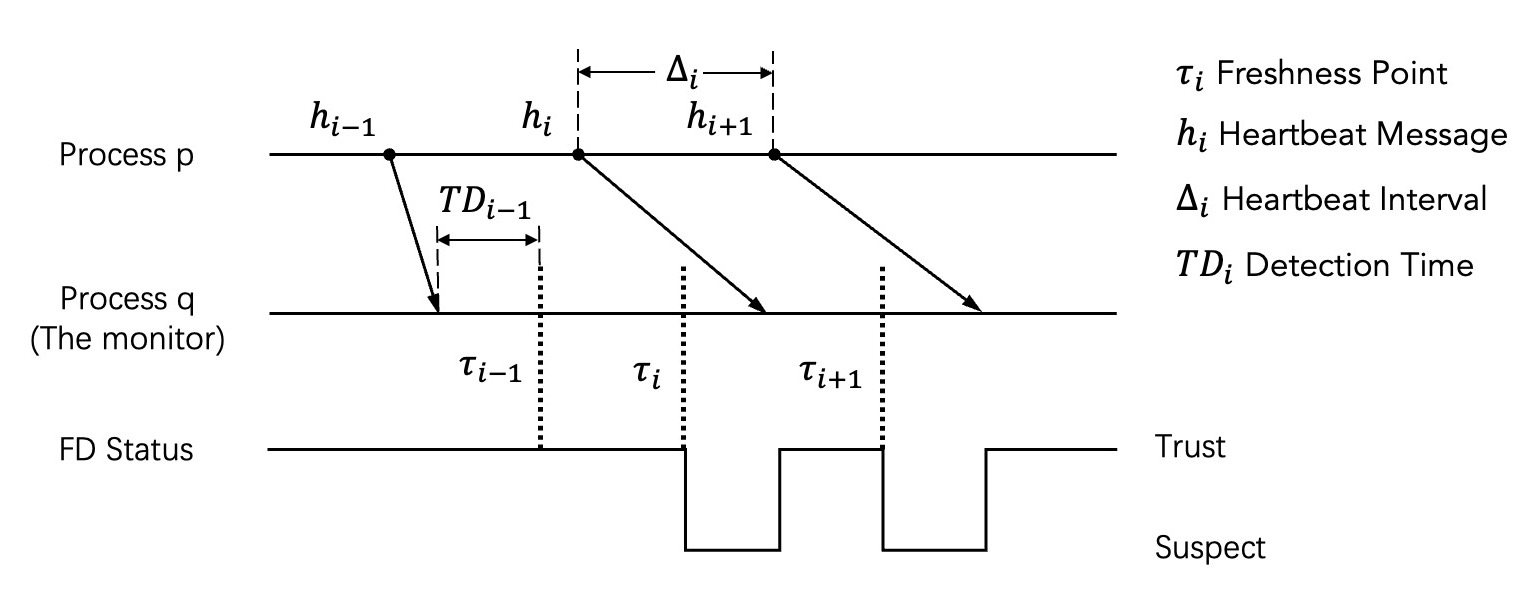}
\vspace*{-5mm}
\caption{Estimating the arrival time of the next heartbeat to suspect failures.}
\vspace*{-2mm}
\label{fig:fd}
\end{figure}

In our distributed system setup, we consider a classic Push-FD model~\cite{felber_push_pull_1999}, where every node $p$ sends heartbeats to its counterparts $q$ every $\Delta_i$ milliseconds (Figure~\ref{fig:fd}). To assess the status of $p$ ($trust$ or $suspect$), $q$ computes an estimate of the arrival date, also called as \emph{freshness point}, $\tau_i$ of the next heartbeat $h_i$. If $q$ receives $h_i$ before $\tau_i$, then it considers that $p$ is up. Otherwise, $q$ starts suspecting $p$ of having crashed. The time that elapses between the emission of $h_i$ and $\tau_i$ is the \emph{detection time} $T_{D}$. 

The FD generates a \emph{freshness point} for the monitored node’s next heartbeat based on its previous behaviors. The \emph{freshness point} is composed of two parts, an \emph{estimated arrival} ($EA$) and a \emph{safety margin} ($\alpha$). The function of a \emph{safety margin} is to mitigate the impact of unexpected and erratic delays and thus to reduce the probability of a false positive detection \cite{chen_toueg_aguilera_2002}. 

We assume that our network follows a crash-stop model, where when a node fails, it will not come back alive. However, it can easily be extended to a crash-recovery model: The FD will consider a node that recovers as a new node. Restricting the model to crash-stop ensures that all nodes behave normally in the time period where incoming data is recorded. Therefore, the main goal of our algorithm is not to detect behavior anomalies, unlike~\cite{log_anomaly_detection, fd_sci_ds, system_fd_dl}. Instead, our FD aims to learn the pattern of heartbeat signals sent by each node in real-time, and to determine an expected arrival time with a high quality of detection: shortest possible detection time, highest possible accuracy.

Three different metrics are used to assess FD performance. The first and the most significant one is the \emph{probability of availability} ($P_{A}$), which is the ratio of the safe predictions over the total predictions. Making a safe prediction means the arrival time of the next heartbeat is earlier than the predicted time, and this is to avoid false-positive results where a node is wrongly suspected. Second, the \emph{detection time} ($T_{D}$) is the difference between the forecast time stamp and the actual reception time stamp. This metric complements the $P_{A}$: it reduces the gap between the predicted time and the actual time, since a model can achieve a high $P_{A}$ by always predicting absurdly long arrival times. Thus, both metrics balance each other: improving $T_{D}$ aggressively increases the rate of false-positives, while doing so for $P_{A}$ increases the risk of false-negatives. Our last metric is the \emph{computation time} $T_{C}$. This is important because of the generally high computational cost of non-linear machine learning methods. An FD whose next prediction takes longer to compute than the next arrival date is pointless.

\section{Baseline}
\label{baseline}
We use Chen's FD (CFD)~\cite{chen_toueg_aguilera_2002} as our baseline for comparison. CFD predicts the next arrival time (Expected Arrival \textit{EA}) upon receiving a heartbeat message. Equation~\ref{equation1} shows how CFD computes its prediction as a statistical analysis of the $n$ latest reception times. To reduce the ratio of false positives, CFD adds a constant safety margin ($\alpha$) to \textit{EA}.  


\begin{equation}\label{equation1}
  \vspace*{-2mm}
	EA_{k+1} \approx \frac{1}{n}\left(\sum_{i=k-n}^kA_{i} - \Delta_{i}*i \right)+(k+1)*\Delta_{i}
  \vspace*{2mm}
\end{equation}

One of the weaknesses of Chen is the constant $\alpha$. Setting $\alpha$ by default reduces the performance of CFD for highly unstable traces. Besides, selecting the optimal value for $\alpha$ is a delicate task. It depends heavily on the network environment, and requires thorough monitoring prior to the deployment of CFD.

\section{Our ML-Based Failure Detector Design}
\label{architecture}



Similarly to CFD, our approach also bases its prediction on an analysis of the $\eta$ latest receptions. We use the long short-term memory (LSTM) model: it can effectively retain important long-term information, and it can quickly fit the data compared to traditional time series forecasting models~\cite{lstm_time}. Moreover, its fitting method is non-linear, which makes its fitting potential stronger than ordinary linear models. 

\begin{figure}[h]
  \centering
  \vspace*{-2mm}
  \includegraphics[scale=0.24]{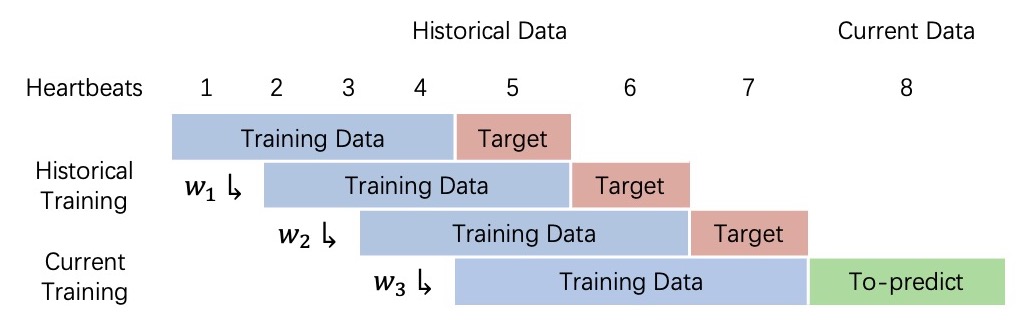}
  \vspace*{-2mm}
  \caption{Our real-time non-accumulative LSTM construction mechanism}
  \label{fig:RTNA_LSTM}
\end{figure}

As our goal is to make real-time predictions, we train our LSTM model real-time and non-accumulatively: we only keep the $\eta$ most recent data items as the training set, and always feed the model with the nearest fixed-size of data. Although we abandon data that falls out of the range, their impact are kept in the parameters. For each real-time training, the model retains the memory of the last training results, and will not restart from scratch. Since we don't store historical data for future training, their impact gradually decreases over time. This training method has two main advantages. First, it speeds up the training process and reduces its computational cost by retaining the memory of the last training results. Second, it gradually reduces the impact of the least recent historical data; this is essential to accommodate the high volatility of network link behaviours. Figure \ref{fig:RTNA_LSTM} illustrates the mechanism of our model.


We need to customize the adjustment of the prediction of our LSTM algorithm, because loss functions such as \emph{mean squared error} treat negative and positive errors equally. This cannot do for an FD, whose objective is to minimize false positives. Our solution is to write our own loss function: its goal is to tilt our FD's output towards over-estimations of the next arrival date. For this purpose, we design a new loss function that adds a multiplier to the results whenever a negative prediction occurs. By using such a loss function, our model achieves a 95\% accuracy rate on its predictions. 

To improve the $PA$ further, we add a dynamic \emph{safety margin} based on the ${\epsilon}$ most recent errors, unlike our baseline model which uses a static one. A \emph{safety margin} is used to help the model adapt to any unexpected and unstable delay when it occurs, and such a delay is unbounded and hard to detect. So the goal is to adapt the model as soon as such a series of delays occur, so having a dynamic array that keeps the ${\epsilon}$ most recent errors enables the model to adjust in time from the latest error. Altogether, our model computes its next freshness point $\tau$ with Equation \ref{equation2}.


\vspace*{-5mm}
\begin{equation}\label{equation2}
	\tau_{k+1} \approx LSTM.predict(k-\eta) + \frac{1}{\epsilon}\sum_{i=k-\epsilon}^{k}error_i
\end{equation}
\vspace*{-3mm}

In the LSTM model, three parameters are decided by exhaustive search: $\eta$ (training data set size), the batch size, and the epoch number. The training data set size $\eta$ decides how much data to learn each time while real-time training; the batch sizes and epoch numbers influence the accuracy and timeliness of training from the structure and nature of training. Grid search is used to search for best combinations. Table \ref{tab:table1} shows the top ten combinations of $P_{A}$ among all since $P_{A}$ is the most critical property for evaluation. The top three combinations share the same $P_{A}$ and similar $T_{D}$, however, the second one has an outstanding $T_{C}$ than the other two. As a result, we used a $\eta$ of 500, a batch size of 64 and an epoch of 5.


\begin{table}[h]
\centering
\begin{tabular}{cccccc}
\hline
\textit{$\eta$}  & \textit{batch size} & \textit{epoch} & $P_{A}$     & $T_{D}$ & $T_{C}$\\
\hline
500          & 32         & 5      & 0.9957 & 9.4416                      & 72.9630          \\
500         & 64         & 5      & 0.9957 & 9.7832                      & 48.1018          \\
500         & 64         & 10     & 0.9957 & 9.1369                      & 74.2897          \\
1000        & 32         & 10     & 0.9956 & 11.1584                     & 150.0853         \\
1000        & 64         & 10     & 0.9947 & 12.2609                     & 89.4486          \\
500         & 32         & 10     & 0.9945 & 9.2335                      & 116.3436         \\
100         & 32         & 5      & 0.9941 & 8.0131                      & 45.8082          \\
1000        & 32         & 5      & 0.9938 & 11.3694                     & 83.6161          \\
1000        & 64         & 5      & 0.9938 & 11.4039                     & 53.5777          \\
100         & 32         & 10     & 0.9931 & 7.5950                      & 61.3970        \\
\hline
\end{tabular}
\vspace*{2mm}
\caption{Parameter combinations that achieve top-10 $P_{A}$ performance}
\label{tab:table1}
\end{table}

 \vspace*{-8mm}

\section{Performance evaluation}
\label{results}

We assess our approach on top of real traces: heartbeat transmission logs collected over a week from 9 nodes on the PlanetLab network (http://www.planet-lab.org/). Nodes $1$ through $9$ generate a heartbeat message containing the sender ID and a sequence number every 100 milliseconds, and send it to node $0$, which we consider as the monitoring node. Upon reception of a heartbeat, node $0$ records its arrival time in the log associated with the sender.

We set the parameter values of Equations~\ref{equation1} and~\ref{equation2} as follows: the heartbeat arrivals window size $n$ for Chen's FD (CFD) is 1000; for our Machine Learning-based FD (MLFD), the heartbeat window size $\eta$ is 500, and the error window size $\epsilon$ is 10. We aim to compare the performance of MLFD with that of CFD. However, CFD's performance depends highly on the value of its constant \emph{safety margin} $\alpha$. To allow for a fair comparison, we align the $P_{A}$ values of both models before comparing their $T_{D}$; this leads to a value of 680ms for $\alpha$. 


\begin{figure}[h]
  \centering
  \includegraphics[scale=0.4]{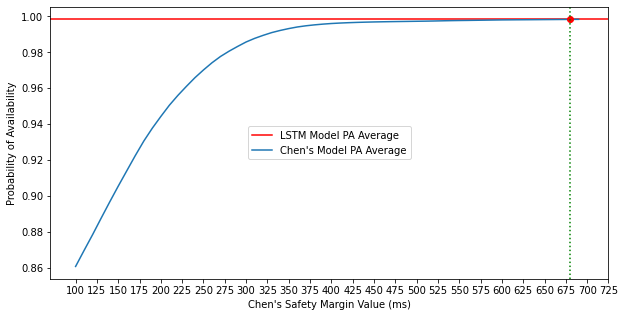}
  \vspace*{-8mm}
  \caption{Influence of CFD's safety margin on its accuracy.}
  \label{fig:comp}
  \vspace*{-2mm}
\end{figure}

Figure~\ref{fig:comp} shows the alignment mechanism after running both FDs on all 9 links: the blue curve shows the correlation between $P_{A}$ and $\alpha$ for CFD, and the red line is the average $P_{A}$ for MLFD (approximately 99.84\%). The plot shows that, for CFD's $P_{A}$ to exceed 99.84\%, its $\alpha$ must be above 680ms. 

Table \ref{tab:table4} gives the comparison of CFD and MLFD's $P_{A}$ and Table \ref{tab:table5} compares results in terms of \emph{detection time} $T_{D}$. These results show how much the \emph{safety margin} of CFD affects its \emph{detection time}. For a similar $P_{A}$ outcome, MLFD detects failures much faster than CFD on any given link. There is a caveat to these excellent results: MLFD consumes a lot of CPU to compute its predictions. In the last column, we add an extra metric: the average \emph{computation times} $T_{C}$ for MLFD's next prediction. We obtained these results on Linux CentOS v6.5 running on 2 cores of an Intel Xeon E5 2.2GHz with 128GB of memory.


We don't include \emph{computation time} for CFD because they remain way below the \emph{detection time}, and are therefore insignificant. As the table shows, MLFD's $T_{C}$ will delay the $T_{D}$ in most cases. The non-linear characteristics of the calculation and the algorithm’s complexity make the computation time longer than its calculated prediction interval, and $T_{C}$ depends highly on the computation power of the server, it might vary greatly on different chips.


\begin{table}[h]
\centering
\begin{tabular}{cccc}
\hline
Link    & Chen's $P_{A}$ (680ms) & MLFD's $P_{A}$         & Chen's $P_{A}$ (690ms) \\ \hline
1       & 1                 & 0.999127 & 1                 \\
2       & 1                 & 0.997104 & 1                 \\
3       & 1                 & 0.998439 & 1                 \\
4       & 1                 & 0.996781 & 1                 \\
5       & 1                 & 0.999165 & 1                 \\
6       & 1                 & 0.998457 & 1                 \\
7       & 0.989449 & 0.998765 & 0.989559 \\
8       & 0.995549 & 0.998366 & 0.995732 \\
9       & 1                 & 0.999055 & 1                 \\
Average & 0.998333 & 0.998362 & 0.998365 \\ \hline
\end{tabular}
\vspace*{2mm}
\caption{$P_{A}$ comparison between CFD and MLFD}
\label{tab:table4}
\end{table}
\vspace*{-2mm}

\begin{table*}[h!]
\centering
\begin{center}
\begin{tabular}{clllll}
\hline
Link    & \multicolumn{1}{c}{CFD $T_{D}$ (680ms)} & \multicolumn{1}{c}{MLFD $T_{D}$} & \multicolumn{1}{c}{CFD $T_{D}$ (690ms)} & \multicolumn{1}{c}{MLFD $T_{C}$}  \\ \hline
1       & 679.093                          & 12.926                   & 689.093                  & 123.847                          \\
2       & 510.860                          & 96.087                   & 520.860                  & 124.529                          \\
3       & 676.351                          & 21.827                   & 686.351                  & 124.682                          \\
4       & 584.397                          & 148.897                  & 594.397                  & 124.836                          \\
5       & 680.019                          & 12.410                   & 690.019                  & 125.254                          \\
6       & 678.190                          & 20.400                   & 688.190                  & 125.199                         \\
7       & 672.419                          & 26.823                   & 682.343                  & 125.754                         \\
8       & 667.905                          & 32.072                   & 677.781                  & 126.229                          \\
9       & 675.992                          & 16.160                   & 685.992                  & 126.670                          \\
Average & 647.247                          & 43.067                   & 657.225                  & 125.222                          \\ \hline
\end{tabular}
\vspace*{3mm}
\caption{$T_{D}$ comparison between CFD and MLFD}
\label{tab:table5}
\end{center}
\end{table*}
\vspace*{-8mm}

\section{Discussion}
\label{discussion}

At the beginning of the experiment, we were concerned that the data fluctuations caused by the delays are arbitrary and non-regular, which might prevent the MLFD from simulating the prediction curve well. But the excellent results show that MLFD can handle data sets that do not present strong regularity. We think the main reason is that, when a significant delay occurs, the following data delays will show a certain degree of consistency. Although MLFD cannot predict a sudden and major delay variation, it can sensitively capture the change and adjust the model parameters to adapt to the delays if they occur in bursts. And once the burst ends, the MLFD will adapt back to the normal trend.

However, the heavy calculation cost does indeed weigh on the performance of our MLFD. On average, the $T_{C}$ of the MLFD is three times as long as its $T_{D}$, which seriously affects the performance of the MLFD. This is obviously a major challenge for the application of machine learning in the field of failure detection. Nevertheless, we remain optimistic about our approach. One reason is that, despite the computational cost issue, the accuracy and \emph{detection time} of our MLFD is still significantly better than that of CFD. The other reason is that the processing power of the server at our disposal to run the MLFD is relatively low. Running our model on a more powerful server would shorten the calculation time significantly.

\section{Conclusion}
\label{conclusion}
In this paper, we present a preliminary study about the feasibility of failure detector implementations based on machine learning algorithms. Our results suggest that ML may be a viable approach: our LSTM-based FD implements a unilateral penalty loss function, dynamic \emph{safety margin}, and it trains on the 500 most recent message receptions in real-time. Upon comparison with Chen’s, our FD achieves much shorter \emph{detection times} for a similar \emph{probability of availability}, but at a significant computation cost. 

Our results illustrate the potential of machine learning models in this field, and we hope it will elicit further research. Our next step is to refine our prototype, and to test it against more aggressive FDs with dynamic \emph{safety margins}~\cite{MYBELOVED_PROFESSORS_PAPER_LETS_CITE_IT}. We also intend to pursue our work with a focus on the trade-off between \emph{probability of availability} and \emph{computation time}. Evolved machine learning models, such as the transformer’s attention model, seem like strong candidates for better performance. The advent of DPUs~\cite{dpu} also opens another promising avenue of research for our work: FDs are obvious candidates for network function virtualization, and SmartNICs have the power to support advanced implementations such as our MLFD.

\bibliographystyle{IEEEtran}
\bibliography{references.bib}

\end{document}